\documentclass[letter]{aa}

\usepackage{graphicx}
\usepackage{xcolor}
\usepackage{hyperref}
\usepackage{txfonts}
\usepackage{booktabs}
\usepackage{threeparttable}
\usepackage{etoolbox}

\hypersetup{colorlinks=true,citecolor=blue,linkcolor=blue,urlcolor=blue}

\newcommand{\uJy}{\ensuremath{\mu\mathrm{Jy}}}

\newcommand{\rev}[1]{#1}

\begin{document}
\title{A radio counterpart of the local little-red-dot
J092359+415736}
\titlerunning{The radio counterpart of the local LRD J092359+415736}

\author{T. An\inst{1}
\and Ruqiu Lin\inst{2}
\and Ailing Wang\inst{3}}

\institute{Department of Astronomy, University of Science and Technology of China, 96 Jinzhai Road, Hefei, Anhui 230026, China
\email{antao2008@ustc.edu.cn}
\and
Department of Astronomy, University of Massachusetts Amherst, Amherst, MA 01003, USA \email{ruqiulin@umass.edu}
\and
Yunnan Observatories, Chinese Academy of Sciences, Kunming 650216, China \email{wangailing@ynao.ac.cn} 
}
\date{Received; accepted}

\abstract
{Little red dots (LRDs) are compact red sources at $z\gtrsim4$. Broad lines suggest accreting black holes, yet radio and X-ray measurements are usually weak. Local LRDs permit multi-frequency, multi-epoch radio studies inaccessible at high redshift.}
{We aim to determine whether archival radio images contain a counterpart to the local LRD J092359+415736 at $z\approx0.323$, to quantify its long-term radio variability, and to identify the plausible physical origin of the radio emission.}
{\rev{We measure the flux density at the optical position in FIRST, LoTSS, RACS, VAST, and VLASS products covering 0.144--3.0 GHz and 1994--2024, and assess long-term variability across the three-decade baseline.}}
{\rev{A compact 144 MHz counterpart is detected at $42\sigma$ in LoTSS, coincident with the \textit{Gaia} position, and is recovered in FIRST, RACS, VAST, and VLASS. The LoTSS and ASKAP measurements give $\alpha\simeq-0.30\pm0.11$. The source exhibits multi-timescale radio variability: FIRST at 1.4~GHz (1994) is a factor of $\sim2.6$ below the ASKAP measurements at similar frequencies (2021--2024) and below the extrapolated LoTSS--ASKAP spectrum, indicating long-term brightening over 26~yr, while the 3~GHz flux density rises by $2.3\sigma$ between VLASS epochs 2.2 and 3.2, suggesting shorter-term variability}. A star-formation-only origin would require a rate of $\simeq200\,M_\odot\,\mathrm{yr}^{-1}$, compared with an unobscured H$\alpha$ upper limit of $3.3\,M_\odot\,\mathrm{yr}^{-1}$, and hence an extreme, heavily obscured starburst. Ordinary single radio supernovae and tidal disruption events are strongly disfavoured by the luminosity and 30-yr recovery history, while refractive interstellar scintillation, if responsible, itself requires an ultracompact nuclear emitter.}
{The persistent radio counterpart of J092359+415736 is best explained by accretion onto its central black hole. \rev{Among the three radio-detected local LRDs, it is intermediate in luminosity and the only one with reported multi-year variability, adding a temporal dimension to the emerging radio diversity of the class.}}


\maketitle
\nolinenumbers
\let\linenumbers\relax

\section{Introduction}
\label{sec:introduction}

The James Webb Space Telescope (JWST) has revealed large numbers of red, compact objects with blue rest-frame ultraviolet continua and broad Balmer emission at $z\gtrsim4$ \citep[e.g.,][]{2024ApJ...963..129M,2024ApJ...964...39G}.
These ``Little Red Dots'' (LRDs) cannot be explained by unobscured quasars or ordinary stellar populations alone. Their broad Balmer lines indicate accretion in at least some LRDs and provide virial black-hole mass estimates. Frequent X-ray non-detections and the low radio detection rate nevertheless remain central physical puzzles \citep[e.g.][]{2024ApJ...976...96P,2025A&A...693L...2P, 2026A&A...706A.372M,2025ApJ...991...37A}.

Radio observations can help determine whether the low detection rate reflects intrinsically weak nuclear emission, absorption by dense gas, or the absence of extended synchrotron structures. Star formation produces spatially extended, slowly evolving synchrotron emission from cosmic-ray electrons accelerated in past supernova remnants. A compact jet, corona, or nuclear wind may instead be variable or self-absorbed. \rev{Free--free absorption and CMB inverse-Compton losses may further suppress the detected emission \citep{2014MNRAS.438.2694G}.} Most high-redshift LRDs are too faint for these possibilities to be separated. Individual searches and stacking analyses of LRDs and related JWST-selected broad-line AGN have mostly yielded upper limits, with only rare radio detections \citep{2025ApJ...991...37A,2025A&A...693L...2P,2025ApJ...986..130G,2026A&A...706A.372M}.

Low-redshift LRDs can be examined in much greater detail than their high-redshift counterparts. \citet{2025ApJ...980L..34L} identified seven low-redshift objects with V-shaped optical--near-infrared continua, broad lines, and emission-line ratios resembling those of high-redshift LRDs. The first published radio counterpart, J1047+0739~\citep{2026ApJ...997..364L}, is a stable, unresolved, steep-spectrum source \citep{2026A&A...707L..17R}. \rev{\citet{2026arXiv260816200R} have also reported the more radio-luminous local analogue J2048.} Here we analyse J092359+415736, the member of the \citet{2025ApJ...980L..34L} sample covered by the Faint Images of the Radio Sky at Twenty centimetres (FIRST, \citealt{1995ApJ...450..559B}) survey, \rev{the LOFAR Two-metre Sky Survey} (LoTSS, \citealt{2017A&A...598A.104S,2026A&A...707A.198S}), the Rapid Australian Square Kilometre Array Pathfinder (ASKAP, \citealt{2021PASA...38....9H}) Continuum Survey (RACS, \citealt{2021PASA...38...58H}), \rev{the Variables And Slow Transients survey (VAST, \citealt{2013PASA...30....6M})}, and the Very Large Array Sky Survey (VLASS, \citealt{2020PASP..132c5001L}), enabling a uniform multi-frequency analysis spanning 0.144--3.0 GHz and three decades.

\section{Source properties}
\label{sec:data}
 
J092359+415736 has the optical position of $09^{\rm h}23^{\rm m}59.8989^{\rm s}$, $+41^\circ57\arcmin36.6909\arcsec$, with uncertainties of $\sigma_{R.A.}=1.2\rm \, mas$ and $\sigma_{Dec.}=0.8\rm \, mas$, based on the \textit{Gaia} DR3 astrometric catalog. Sloan Digital Sky Survey (SDSS) spectroscopy \citep{2026AJ....171...52K} gives redshift $z=0.323$ and classifies the source as a broad-line AGN. The published optical analysis gives $\log(M_{\rm BH}/M_\odot)=6.86$ (single-epoch virial systematic $\approx0.5$ dex) and $\log(M_\star/M_\odot)=10.03\pm0.20$, together with a V-shaped continuum and AGN-dominated emission-line ratios on the Baldwin--Phillips--Terlevich (BPT) diagram \citep{1981PASP...93....5B,2025ApJ...980L..34L}. 
The optical-to-near-infrared spectral energy distribution (SED) can be approximated phenomenologically by two blackbody components, with the hotter component at $T\simeq4600$ K (Appendix \ref{sec:sed}). This characteristic temperature is consistent with LRD models in which radiation from an accreting black hole is thermalised and reprocessed by an optically thick, dense gas envelope \citep[e.g.][]{2025arXiv251121820D,2026ApJ..1000...90I,2026ApJ...997..364L}.
\rev{The source is detected in all four \textit{WISE} bands, and the published mid-infrared colours are also reproduced by a starburst template (Appendix~\ref{sec:sed}).}
Its published broad-H$\alpha$ luminosity, $(6.64\pm0.46)\times10^{41}\,\mathrm{erg\,s^{-1}}$ \citep{2024SCPMA..6709811L}, gives $\lambda L_\lambda(5100\,\text{\AA})=1.7\times10^{43}\,\mathrm{erg\,s^{-1}}$ from the \citet{2005ApJ...630..122G} relation. With $L_{\rm bol}=9.8\lambda L_\lambda(5100\,\text{\AA})$, this corresponds to $L_{\rm bol}\simeq1.6\times10^{44}\,\mathrm{erg\,s^{-1}}$ and $L_{\rm bol}/L_{\rm Edd}\simeq0.18$. 
The 0.5-dex virial-mass systematic alone gives $L_{\rm bol}/L_{\rm Edd}\simeq0.06$--0.58, with additional uncertainty from broad-line decomposition and the bolometric correction.

\section{Flux density measurements and variability} 
\label{sec:methods}

\begin{figure}
\centering
\includegraphics[width=\columnwidth]{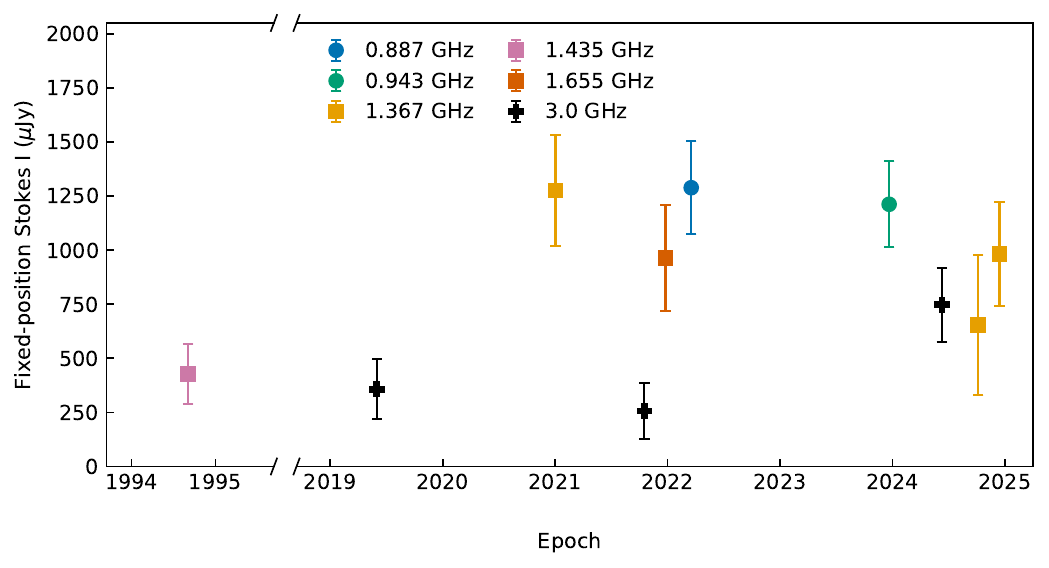}
\caption{\rev{Radio flux densities at the \textit{Gaia} DR3 position in the 10 independent archival epochs. Symbol colour and shape encode the observing frequency. Flux densities are shown at their observed frequencies and are not scaled to a common reference frequency. Error bars show $\sigma_{\rm tot}$.}}
\label{fig:varsummary}
\end{figure}

We analysed one 1.435-GHz FIRST image \citep{2015ApJ...801...26H}, \rev{a 144 MHz LoTSS DR3 mosaic}  \citep{2026A&A...707A.198S}, \rev{RACS snapshots at 0.887, 1.367, and 1.655 GHz} \citep{2024PASA...41....3D,2025PASA...42...38D}, \rev{VAST images at 0.943 and 1.367 GHz} \citep{2013PASA...30....6M}, and three 3-GHz VLASS Quick Look images \citep{2020PASP..132c5001L}. \rev{Figure~\ref{fig:varsummary} shows the flux densities on the 10 independent GHz epochs. Figure~\ref{fig:cutouts} shows the corresponding GHz cutouts. Figure~\ref{fig:lotss} shows the LoTSS DR3 image. } The observing dates, restoring beams, and measurements are listed in Table~\ref{tab:photometry}.

At each epoch we fitted an elliptical Gaussian matched to the restoring beam and fixed at the \textit{Gaia} DR3 position, together with a local planar background. The Gaussian was normalised to unity at its centre, so its fitted scale gives the peak intensity. This peak intensity equals the source flux density when the source is unresolved. At $z=0.323$, the major axes of the LoTSS, FIRST, VLASS, and RACS beams correspond to $\sim28$, $\sim25$, $\sim13$, and up to $\sim180$ kpc, respectively\footnote{We adopt a flat $\Lambda$CDM cosmology with $H_0=70~\mathrm{km\,s^{-1}\,Mpc^{-1}}$ and $\Omega_{\rm m}=0.3$, matching \citet{2025ApJ...980L..34L}. At the source redshift, $1\arcsec=4.69$ kpc. Spectral indices are defined by $S_\nu\propto\nu^\alpha$.}. Positional coincidence at these resolutions does not establish nuclear compactness. The baseline fit used a circular aperture with radius 1.6 times the beam major axis. Repeating the fit with radii of 1.2 and 2.0 times the major axis and with constant or planar backgrounds produced amplitudes consistent within the adopted errors.

We define the empirical image-plane error $\sigma_{\rm stat}$ as the largest of the local background \textit{rms}, the robust scatter of identical fits at blank positions where available, and the scatter among the six fitting variants. These estimates probe overlapping noise contributions, so we take their maximum rather than add them in quadrature. The total error is $\sigma_{\rm tot}^2=\sigma_{\rm stat}^2+(f_{\rm cal}S_\nu)^2$, with $f_{\rm cal}=0.05$, 0.10, and 0.15 for FIRST, \rev{LoTSS/RACS/VAST}, and VLASS, respectively. The conservative 15\% term accounts for the lower calibration fidelity of the VLASS images relative to the $\lesssim3\%$ quoted for VLASS Single Epoch products. Free-centroid fits were used only as an astrometric check, not to select positive noise peaks. Their offsets from the \textit{Gaia} position are $0.25\arcsec$ for LoTSS, $<0.8\arcsec$ for FIRST and VLASS, and $4\arcsec$--$6\arcsec$ for the elongated RACS beams, the latter being $\lesssim0.5$ of the beam minor axis. Fixing the centroid changes the VLASS amplitudes by $<20\%$, within the adopted errors.
\rev{Allowing small positional offsets changes the ASKAP amplitudes by $\lesssim8\%$ for detections above $3\sigma$, and FIRST/VLASS residual maps show no recurring compact component at a common offset (Appendix~\ref{app:offsetfit}). The survey resolutions do not separate nuclear emission from unresolved host-galaxy emission.}
For FIRST we use the raw image-plane amplitude, $428\pm139\,\uJy$, and source-injection tests confirm no measurable bias from the fixed-position fit (Appendix~\ref{app:offsetfit}). The FIRST measurement is a factor of $\sim2.6$ below the LoTSS--ASKAP spectrum (Appendix~\ref{app:racsinternal}).

\rev{LoTSS DR3 recovers a compact source at the \textit{Gaia} position with a peak of $2.13\,\mathrm{mJy}\,\mathrm{beam}^{-1}$ ($42\sigma$ relative to the local \textit{rms} of $51\,\uJy\,\mathrm{beam}^{-1}$) and a fitted flux density of $2234\pm235\,\uJy$ (Appendix~\ref{app:lotss}). The peak lies $0.9\arcsec$ from the \textit{Gaia} position. No other compact source exceeds $4\sigma$ within $60\arcsec$ (Fig. \ref{fig:lotss}), so the ASKAP measurements are not blends of a second component.} Six ASKAP/RACS and VAST measurements during 2021--2024 range from $0.65$ to $1.29$ mJy, with five detected above $3\sigma$, confirming persistent emission at GHz frequencies. 
\rev{Although the LoTSS and ASKAP data are not strictly simultaneous, their flux densities follow a single power law: RACS-low at $0.887\,\mathrm{GHz}$ gives $\alpha=-0.30\pm0.11$ relative to LoTSS, and the remaining ASKAP bands at $0.94$--$1.66\,\mathrm{GHz}$ lie within $\sim12$ per cent of that spectrum. This spectrum predicts $\sim 1.1$~mJy at $1.435\,\mathrm{GHz}$. FIRST measurement in 1994, with a $5.4\arcsec$ beam closely matched to the LoTSS $6.0\arcsec$ beam, is a factor of $\sim2.6$ fainter (Appendix~\ref{app:racsinternal}). The FIRST deficit is not an artefact of missing extended emission, since LoTSS already recovers $\sim2.2\,\mathrm{mJy}$ at the same angular scale.} The 26~yr FIRST-to-ASKAP contrast constitutes the main long-baseline variability evidence. Between VLASS epochs 2.2 (2021 October) and 3.2 (2024 June), the 3~GHz flux density rises from $257\pm129$ to $748\pm170\,\uJy$, a factor of $\sim3$ increase over 2.6~yr ($2.3\sigma$).

\begin{table}
\centering
\caption{Radio flux densities of J092359+415736.}
\label{tab:photometry}
\footnotesize
\setlength{\tabcolsep}{3.2pt}
\begin{tabular}{@{}lcccc@{}}
\toprule
Data set & Date & $\nu$ & Beam & $S_\nu$ \\
 & & (GHz) & (arcsec) & (\uJy) \\
\midrule
\rev{LoTSS DR3} & \rev{2014--2024$^{\mathrm{a}}$} & \rev{0.144} & \rev{$6.0\times6.0$} & \rev{$2234\pm235$} \\
FIRST & 1994-09-04 & 1.435 & $5.4\times5.4$ & $428\pm139$ \\
VLASS 1.2 & 2019-05-31 & 3.000 & $2.72\times2.25$ & $357\pm139$ \\
RACS-mid & 2021-01-01 & 1.367 & $39.3\times12.0$ & $1276\pm257$ \\
VLASS 2.2 & 2021-10-17 & 3.000 & $2.76\times2.24$ & $257\pm129$ \\
RACS-high & 2021-12-24 & 1.655 & $38.6\times7.9$ & $965\pm246$ \\
\rev{RACS-low} & \rev{2022-03-18} & \rev{0.887} & \rev{$45.5\times12.2$} & \rev{$1288\pm214$} \\
\rev{VAST-low} & \rev{2023-12-21} & \rev{0.943} & \rev{$36.3\times11.3$} & \rev{$1212\pm198$} \\
VLASS 3.2 & 2024-06-09 & 3.000 & $2.68\times2.11$ & $748\pm170$ \\
\rev{VAST-mid 1} & \rev{2024-10-06} & \rev{1.367} & \rev{$28.9\times8.7$} & \rev{$654\pm325$} \\
\rev{VAST-mid 2} & \rev{2024-12-13} & \rev{1.367} & \rev{$26.9\times7.9$} & \rev{$983\pm241$} \\
\bottomrule
\end{tabular}
\tablefoot{\rev{VAST-low is an inverse-variance stack and is treated as a single epoch. Uncertainties are $\sigma_{\rm tot}$ (Sect.~\ref{sec:methods}).}
\tablefoottext{a}{\rev{LoTSS DR3 mosaic of 2014--2024 pointings. See Appendix~\ref{app:lotss}.}}}
\end{table}

\section{Physical origin of the radio emission}
\label{sec:discussion}

\subsection{Continuous star formation}

For a \citet{2001MNRAS.322..231K} initial mass function (IMF), the global 1.4-GHz star formation rate (SFR) calibration of \citet{2011ApJ...737...67M} gives $\mathrm{SFR}\,[M_\odot\,\mathrm{yr}^{-1}]=6.35\times10^{-29}L_{1.4}\,[\mathrm{erg\,s^{-1}\,Hz^{-1}}]$. \rev{Scaling the LoTSS flux density to 1.4 GHz with the measured $\alpha=-0.30\pm0.11$ gives $L_{1.4}=(3.18\pm0.34)\times10^{30}\,\mathrm{erg\,s^{-1}\,Hz^{-1}}$ and a radio-equivalent SFR of $202\pm22\,M_\odot\,\mathrm{yr}^{-1}$. The quoted errors propagate the radio measurement uncertainties but do not represent the intrinsic scatter of the radio--SFR calibration. } Because the emission-line ratios are AGN-dominated, assigning all narrow H$\alpha$ luminosity ($L_{{\rm H}\alpha,n}=(6.22\pm0.19)\times10^{41}\,\mathrm{erg\,s^{-1}}$) to H\,{\sc ii} regions gives a conservative upper limit on the unobscured SFR. The H$\alpha$ calibration of \citet{2012ARA&A..50..531K} gives $\mathrm{SFR}({\rm H}\alpha)=3.34\pm0.10\,M_\odot\,\mathrm{yr}^{-1}$. Reconciling this value with the radio SFR would require H$\alpha$ attenuation of $\sim4.4$ mag if both tracers arise from the same star-forming population. The radio-inferred rate also corresponds to a specific star formation rate of $\sim19\,\mathrm{Gyr}^{-1}$, more than an order of magnitude above the star-forming main sequence at $z\simeq0.3$ for the measured stellar mass \citep{2014ApJS..214...15S}. Sustaining the observed radio flux density by continuous star formation alone would therefore require an extreme, heavily obscured starburst.

\subsection{Radio supernovae and tidal disruption events}

The rest-frame 1.4-GHz luminosity of J092359+415736 exceeds the upper end of the $10^{26}$--$10^{29}\,\mathrm{erg\,s^{-1}\,Hz^{-1}}$ range observed for luminous late-time radio supernovae \citep{2021ApJ...923L..24S} by more than an order of magnitude. Some radio supernovae remain detectable for decades, but the combination of this extreme luminosity and independent recoveries in 1994, in LoTSS DR3, and in 2019--2024 is unlike known ordinary events, and multiple explosions producing recurrent emission would reduce to the high supernova rate already excluded by the continuous star-formation argument. The corresponding rest-frame luminosity $\nu_{\rm e}L_{\nu_{\rm e}}\simeq(2$--$6)\times10^{39}\,\mathrm{erg\,s^{-1}}$ likewise exceeds the characteristic radio luminosities of most non-jetted tidal disruption events (TDEs), although the observed TDE population is broad and includes radio-luminous jetted or delayed-outflow cases \citep{2020SSRv..216...81A,2024ApJ...971..185C}. A single ordinary TDE cannot naturally reproduce radio emission recovered three decades apart, and the optical light curve, though blind to any pre-2018 flare, shows no transient during 2018--2024 \citep{2026ApJ..1005..150L}. Accounting for both the multi-epoch radio recurrence and the absence of a recent optical counterpart would require a rare and finely tuned scenario, such as a jetted TDE with atypical delayed radio brightening or a long-lived circumnuclear interaction, rather than the ordinary TDE population.

\subsection{Refractive interstellar scintillation}

Refractive interstellar scintillation could contribute to the differences between widely separated survey epochs if a substantial fraction of the radio flux density arises from an ultracompact non-thermal radio component. Appendix~\ref{app:riss} shows that reproducing the observed 3-GHz modulation amplitude would require an emitting region small enough to imply a brightness temperature characteristic of accretion-powered emission. Attributing the variability to scintillation therefore leads to the same inference as an intrinsic origin: a compact accretion-powered radio component.

\subsection{Accretion-powered radio emission}
\label{sec:accretion}

The broad H$\alpha$ line, AGN-dominated BPT ratios, and virial BH mass establish an active nucleus independently of the radio data \citep{2025ApJ...980L..34L}. Continuous star formation and ordinary radio transients are strongly disfavoured, while scintillation would require an ultracompact nuclear emitter. Together with the radio counterpart recovered over three decades and multi-year flux-density changes seen independently at 3 GHz and across the 1.3--1.6 GHz measurements, these results identify black-hole accretion as the dominant power source of the radio emission. The weighted 3-GHz flux density corresponds to $L_{\nu_{\rm e}}(3.97\,\mathrm{GHz})=(1.06\pm0.21)\times10^{30}\,\mathrm{erg\,s^{-1}\,Hz^{-1}}$ and $L_{\rm R}\equiv\nu_{\rm e}L_{\nu_{\rm e}}=(4.2\pm0.9)\times10^{39}\,\mathrm{erg\,s^{-1}}$. Combining this with $L_{\rm bol}\simeq1.6\times10^{44}\,\mathrm{erg\,s^{-1}}$ gives $L_{\rm R}/L_{\rm bol}\simeq2.6\times10^{-5}$, directly quantifying the radio output associated with the radiative accretion power.

A compact jet, a magnetised corona, and shocks driven by a nuclear wind can all produce unresolved radio emission in radio-quiet AGN \citep{2008MNRAS.390..847L,2019NatAs...3..387P}. Within the accretion-powered picture established above, a compact non-thermal outflow provides the most natural explanation for the radio luminosity, long-lived recurrence, and variability. A weak relativistic jet may therefore contribute substantially to the detected emission, although the present data do not distinguish it from coronal or wind-driven synchrotron emission. A quasi-simultaneous broadband radio spectrum, the flux fraction and morphology recovered with Very Long Baseline Interferometry (VLBI), the intrinsic brightness temperature, and any component motion can identify the dominant channel \citep{2023MNRAS.525.6064W}.

\section{Radio diversity and high-redshift non-detections}
\label{sec:implications}

\rev{Before the present work, radio detections had been reported for J1047+0739 and J2048.} Table~\ref{tab:local} compares the three radio-detected local LRDs in redshift, rest-frame 6 GHz luminosity, spectral index where measured, and reported multi-year variability.
J1047+0739 is an order of magnitude fainter than J092359+415736, has a measured steep spectrum, and is unresolved on a scale of $<700$ pc, with no reported variability from 2010 to 2018 \citep{2026A&A...707L..17R}.
\rev{J2048 is a factor of $\sim6$ more luminous, with a flatter published spectrum and no significant change over eight years of VLASS imaging \citep{2026arXiv260816200R}.}
J092359+415736 occupies the intermediate luminosity range, has a LoTSS--ASKAP index close to that of J2048, and is the only one of the three with independent multi-year flux-density changes at 3 GHz and near 1.4 GHz.

The spectral indices span more than 0.5 in $\alpha$. The steep index of J1047+0739 ($\alpha=-0.85$) is consistent with an aged synchrotron population, possibly in a more extended structure or with a stronger magnetic field, and its stability over eight years supports quasi-steady emission. The flatter spectral indices of J092359+415736 and J2048 ($\alpha\simeq-0.30$ to $-0.39$) are more typical of compact, actively replenished synchrotron emission from a jet base or magnetised corona \citep{2008MNRAS.390..847L}. J2048 shares a similar spectral index but is a factor of $\sim6$ more luminous and shows no variability, suggesting that spectral shape alone does not predict variability; the difference may instead reflect Doppler factors, jet inclinations, or accretion states. The emerging picture is one of physical diversity within the local LRD population, analogous to the diversity seen in the broader radio-quiet AGN population at low redshift \citep{2023MNRAS.525.6064W}.

\begin{table}
\centering
\caption{\rev{Radio-detected local LRD analogues.}}
\label{tab:local}
\footnotesize
\setlength{\tabcolsep}{3.5pt}
\begin{tabular}{@{}lccc@{}}
\toprule
 & J1047+0739 & J0923$^{\mathrm{a}}$ & J2048 \\
\midrule
$z$ & 0.168 & 0.323 & 0.433 \\
$L_{6\,\mathrm{GHz}}$ & 0.89 & 21 & 120 \\
$\alpha$ & $-0.85\pm0.24$ & $-0.30\pm0.11$ & $-0.39\pm0.04$ \\
Variability & stable & multi-year & stable \\
\bottomrule
\end{tabular}
\tablefoot{$L_{6\,\mathrm{GHz}}$ is the rest-frame 6 GHz luminosity in $10^{29}\,\mathrm{erg\,s^{-1}\,Hz^{-1}}$, from the 6 GHz detection of J1047+0739 \citep{2026A&A...707L..17R}, the LoTSS measurement of J0923 with $\alpha=-0.30\pm0.11$, and $S_{3\,\mathrm{GHz}}=3.0$ mJy with $\alpha=-0.39$ for J2048 \citep{2026arXiv260816200R}. For J0923, multi-year variability refers to changes on both year and decade timescales (Sect.~\ref{sec:methods}).
\tablefoottext{a}{J092359+415736.}}
\end{table}

\rev{J092359+415736 is a sub-Eddington system ($L_{\rm bol}/L_{\rm Edd}\simeq0.2$), whereas high-redshift LRDs may have different intrinsic accretion rates (Appendix~\ref{app:highz}).} At high redshift, CMB inverse-Compton cooling preferentially suppresses extended, weakly magnetised structures, making compact nuclear components the more promising targets \citep{2014MNRAS.438.2694G,2015MNRAS.452.3457G}. \rev{Placed at $z=5$, the LoTSS flux density of J092359+415736 would fall to a few $\uJy$ at 1.4~GHz} (assuming $\alpha=-0.30$ and no evolution), below the detection threshold of current deep surveys. Its non-detection at high redshift would therefore be entirely consistent with sensitivity limits rather than an intrinsically different radio output. \rev{Within these limits, the persistent and luminous radio emission of J092359+415736 suggests that at least part of the high-redshift population may host similarly bright compact counterparts.} Deep multi-frequency VLBI imaging would offer a direct route to identifying the compact emission.

\section{Conclusions}
\label{sec:conclusions}

\rev{The radio counterpart of J092359+415736 is established by the LoTSS DR3 detection and confirmed across four additional surveys spanning 0.144--3.0 GHz and three decades.} \rev{The combination of radio luminosity, flat spectral index, and multi-timescale variability (a factor of $\sim2.6$ brightening over 26~yr and a tentative $\sim3\times$ rise at 3~GHz over 2.6~yr) identifies accretion-powered emission as the most plausible origin.} With $L_{\rm R}/L_{\rm bol}\simeq2.6\times10^{-5}$, the source falls in the radio-quiet regime, and the short variability timescale constrains the emitting region to light-year scales, consistent with a jet base or magnetised corona.

\rev{J092359+415736 is intermediate in luminosity among the three radio-detected local LRDs and the only one with reported multi-year variability, adding a temporal dimension to the emerging radio diversity of the class.} The three objects together span two orders of magnitude in $L_{6\,\mathrm{GHz}}$ and cover a range of spectral indices and variability behaviours, demonstrating that local LRDs do not constitute a uniform radio population. Whether this diversity reflects intrinsic differences in jet power, accretion rate, or viewing geometry remains to be determined.  Simultaneous broadband spectra, VLBI imaging, and dedicated monitoring of local LRDs can distinguish a weak jet from coronal or wind-driven emission and provide a physical template for interpreting the still largely undetected radio emission of high-redshift LRDs.

\begin{acknowledgements}
The National Radio Astronomy Observatory is a facility of the U.S. National Science Foundation operated under cooperative agreement by Associated Universities, Inc. This publication makes use of data from FIRST, LoTSS, VLASS, RACS, \rev{VAST}, SDSS, and SPHEREx. \rev{LOFAR data products were provided by the LOFAR Surveys Key Science Project (https://lofar-surveys.org/) and were derived from observations with the International LOFAR Telescope.} The Australian SKA Pathfinder (ASKAP) is part of the Australia Telescope National Facility, which is managed by CSIRO. Operation of ASKAP is funded by the Australian Government with support from the National Collaborative Research Infrastructure Strategy. 
Funding for the Sloan Digital Sky Survey (\url{www.sdss.org}) has been provided by the Alfred P. Sloan Foundation, the Heising-Simons Foundation, the U.S. National Science Foundation, and the Participating Institutions. SPHEREx is a joint project of the Jet Propulsion Laboratory and the California Institute of Technology, funded by NASA.
\end{acknowledgements}

\section*{Data availability}

All data used in this study are available from public archives: the radio images from FIRST, \rev{LoTSS}, RACS, \rev{VAST}, and VLASS surveys, and the optical data from SDSS and SPHEREx archives.

\bibliographystyle{aa}
\bibliography{J0923_AA_Letter}

\begin{appendix}

\section{Blackbody-like Optical-NIR SEDs}\label{sec:sed}
\begin{figure*}
    \centering
    \includegraphics[width=0.625\linewidth]{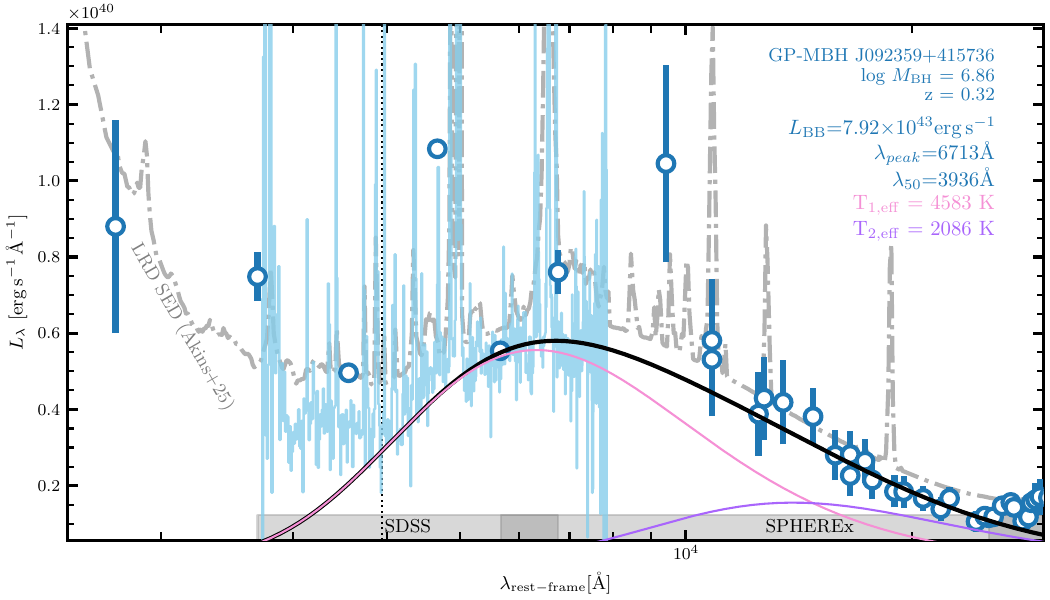}
    \includegraphics[width=0.34\linewidth]{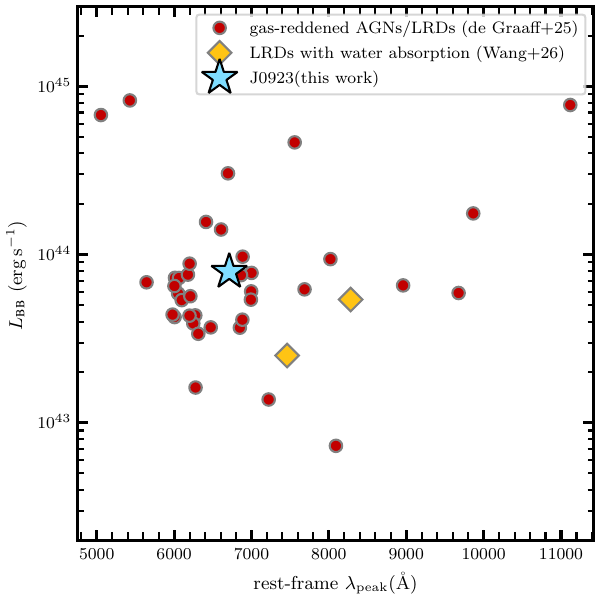}
    \caption{Left: Ultraviolet-Near infrared SED (blue dots) and optical spectrum (blue curve) of J092359+415736. The SED is fitted by two-component blackbody models (black curve) with temperature of $T_{\rm 1, eff} = 4583 \rm\, K$ (pink curve) and $T_{\rm 2, eff} = 2086 \rm\, K$ (purple curve). Right: Integrated blackbody luminosity $L_{\rm BB}$ vs. peak wavelength of blackbody model. The blue star indicates J092359+415736. Red dots and gold diamond show the gas-reddened AGNs/LRDs reported in \cite{2025arXiv251121820D} and \cite{2026arXiv260206024W}. }
    \label{fig:sed}
\end{figure*}

Figure~\ref{fig:sed} shows the ultraviolet-near-infrared of J092359+415736. We fit the SEDs using a two-component blackbody model from 4000 \AA\ to 1.6 $\rm \mu m$, motivated by the results of \cite{2026arXiv260206024W}, who suggested that LRDs may consist of two components with different characteristic temperatures. Unlike the two-component photospheric model adopted by \cite{2026arXiv260206024W}, however, we employ a simplified model using only blackbody spectra. This choice is motivated by the limited sensitivity of the current SDSS spectra and the modest spectral resolution of the SPHEREx data \citep{2026ApJ...999..139B}, which make it difficult to constrain a photospheric model through stellar absorption features. We note that our primary goal is to compare J092359+415736 with other LRDs in terms of their bolometric luminosities and peak wavelengths (which are directly related to the characteristic temperatures). The choice of model has only a minor impact on the determination of these two quantities. As shown in Figure~\ref{fig:sed}, J092359+415736 exhibits a bolometric luminosity and characteristic temperature comparable to those of the majority of known gas-reddened LRDs \citep{2025arXiv251121820D}. \rev{The source is detected in all four \textit{WISE} bands. Its mid-infrared luminosity and colours are also reproduced by the starburst model of \citet{2025ApJ...980L..34L}. The two-blackbody fit is used here only to characterize the optical--near-infrared continuum and its characteristic temperatures. No torus decomposition is attempted.} The integrated blackbody luminosity $L_{\rm BB}$ shown in the right panel of Fig.~\ref{fig:sed} is derived from the model fit and is distinct from the AGN bolometric luminosity $L_{\rm bol}$ derived from the broad H$\alpha$ calibration in Sect.~\ref{sec:data}.

\section{Archival radio cutouts}
\label{app:cutouts}

\begin{figure*}[t]
\centering
\includegraphics[width=\textwidth]{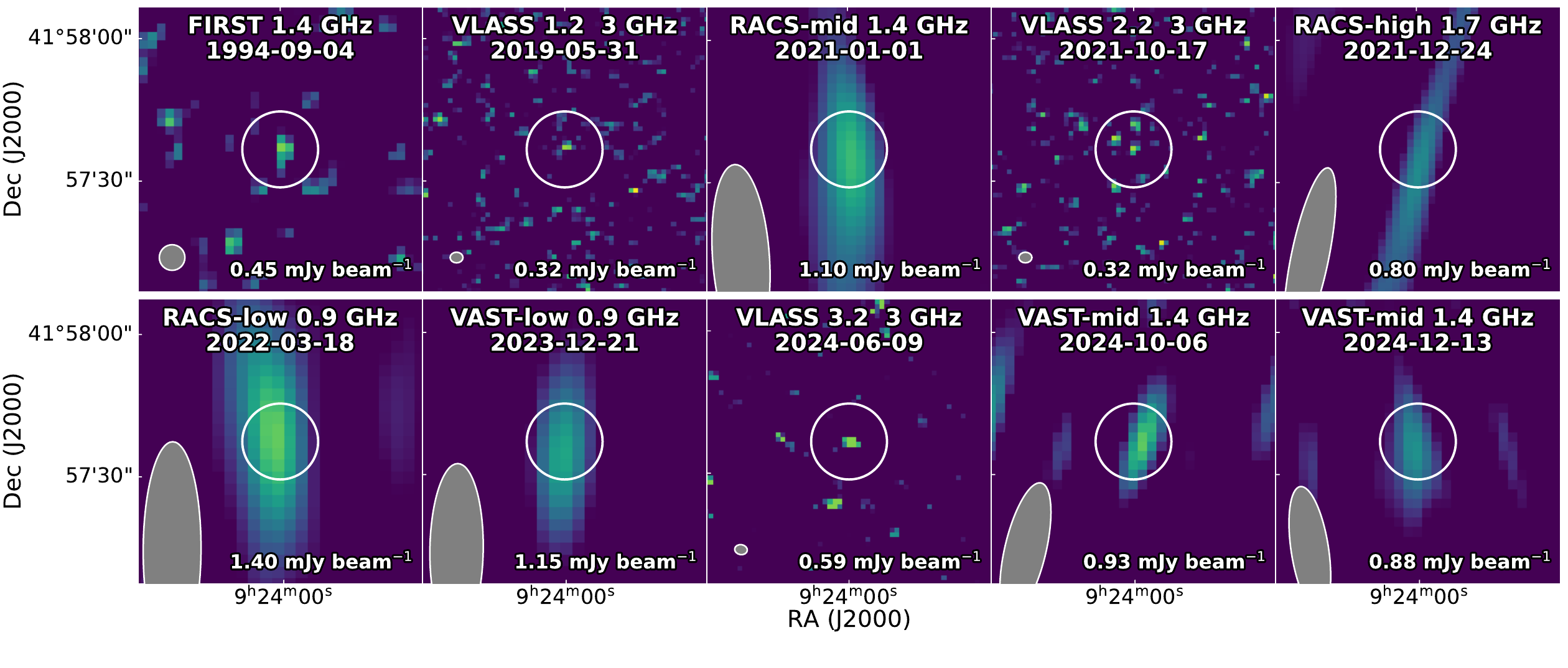}
\caption{Radio cutouts corresponding to the ten GHz measurements in Table~\ref{tab:photometry}, ordered by observing date. Top row: FIRST ($1.435$ GHz), VLASS 1.2 ($3$ GHz), RACS-mid ($1.367$ GHz), VLASS 2.2 ($3$ GHz), and RACS-high ($1.655$ GHz). Bottom row: RACS-low ($0.887$ GHz), the VAST-low stack ($0.943$ GHz), VLASS 3.2 ($3$ GHz), and VAST-mid at $1.367$ GHz on 2024-10-06 and 2024-12-13. The white circle marks the \textit{Gaia} DR3 position used for the forced fits and has radius $8\arcsec$. The number in each panel is the map intensity at that position in $\mathrm{mJy\,beam^{-1}}$, shown for visual reference. These values are not the fitted flux densities listed in Table~\ref{tab:photometry}. Each panel uses an independent linear display range. All panels are $60\arcsec$ on a side. Grey ellipses mark the restoring beams.}
\label{fig:cutouts}
\end{figure*}

\rev{Figure~\ref{fig:cutouts} shows the radio cutouts corresponding to the ten GHz measurements in Table~\ref{tab:photometry}, ordered by observing date and centred on the \textit{Gaia} DR3 position of J092359+415736. The LoTSS DR3 image is shown separately in Fig.~\ref{fig:lotss}.} The white open circle has a radius of $8\arcsec$ and marks the position used for the forced fits. The grey ellipse marks the restoring beam. The labelled number in each panel is the map intensity at that position. Each panel uses an independent linear display range. RACS-mid, \rev{RACS-low, and the VAST-low stack} show a clear excess at the target position, while the low-level stripes in RACS-high are ASKAP snapshot imaging artefacts. The VLASS 1.2 and 2.2 signals are marginal, whereas VLASS 3.2 is more prominent, consistent with the $2.56$, $1.99$, and $4.39\sigma$ forced measurements. \rev{VAST-mid 2 recovers the source at $4.08\sigma$. VAST-mid 1 is a low-significance measurement at $2.01\sigma$.}

\section{LoTSS DR3 144 MHz imaging}
\label{app:lotss}

\rev{The 144 MHz image of J092359+415736 is a $5\arcmin$ cutout from LoTSS Data Release 3 \citep{2026A&A...707A.198S}. The parent product is HEALPix mosaic HP-505, formed from LOFAR High Band Antenna observations processed with {\tt ddf-pipeline} v3.1 \citep{2021A&A...648A...1T}. The mosaic has a restoring beam of $6.0\arcsec\times6.0\arcsec$ and a pixel scale of $1.5\arcsec$. The FITS header does not record an observing date, because DR3 mosaics combine multiple 8 h pointings taken during the 2014--2024 LoTSS campaign. The local rms is $51\,\uJy\,\mathrm{beam}^{-1}$, below the DR3 mosaic median of $92\,\uJy\,\mathrm{beam}^{-1}$, indicating that the image quality at this position is better than the survey average.}

\begin{figure}
\centering
\includegraphics[width=\columnwidth]{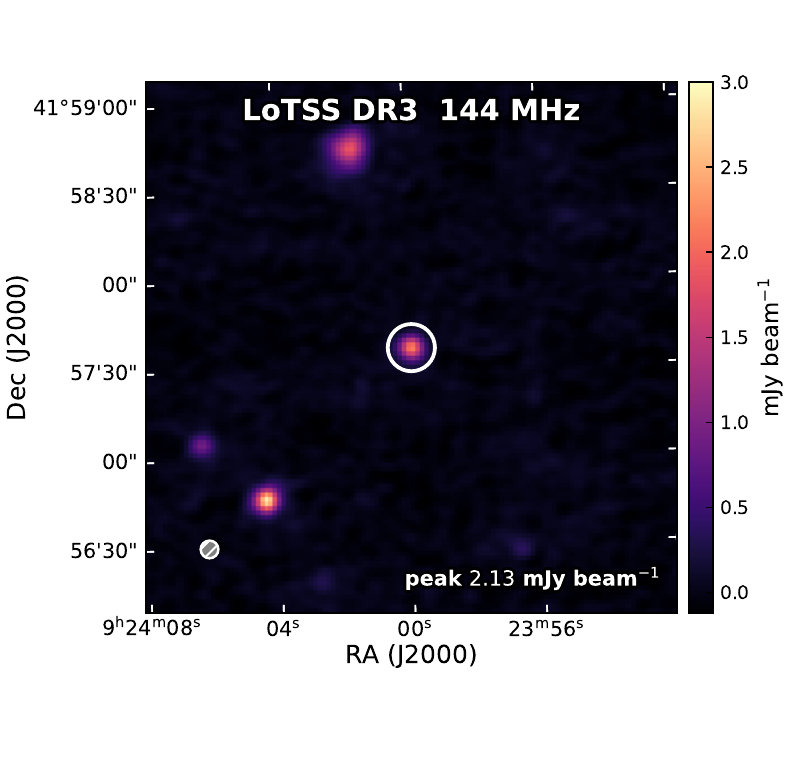}
\caption{\rev{LoTSS DR3 144 MHz image of J092359+415736. The field is $3\arcmin$ on a side. The white circle (radius $8\arcsec$) marks the \textit{Gaia} position used for the fit. The hatched ellipse in the bottom-left corner is the $6.0\arcsec$ restoring beam. The compact source inside the circle has a peak intensity of $2.13\,\mathrm{mJy}\,\mathrm{beam}^{-1}$ ($42\sigma$). The brighter source $71\arcsec$ to the south-east is unrelated.}}
\label{fig:lotss}
\end{figure}

\rev{Figure~\ref{fig:lotss} shows a $3\arcmin$ field centred on the \textit{Gaia} position. No other compact source exceeds $4\sigma$ within $60\arcsec$. We fitted a beam-matched elliptical Gaussian plus a planar background at that position, using the same aperture grid as in Sect.~\ref{sec:methods}. The baseline fit ($r=1.6\,\theta_{\rm maj}$) gives $2234\,\uJy$. Fits with radii $1.2$, $1.6$, and $2.0$ times the beam major axis give $2158$, $2234$, and $2304\,\uJy$, so the aperture scatter is $73\,\uJy$. Combined with a 10 per cent flux-scale term this yields $2234\pm235\,\uJy$ ($S/\sigma_{\rm tot}=9.5$). The DR3 flux-scale accuracy is quoted as 6 per cent (random) and 2 per cent (systematic), so the 10 per cent term is a conservative estimate. The peak intensity is $2.13\,\mathrm{mJy}\,\mathrm{beam}^{-1}$, or 42 times the local rms, and lies $0.9\arcsec$ from \textit{Gaia}. A free-centroid fit is offset by $0.25\arcsec$. A circular Gaussian has FWHM $6.7\arcsec$ against the $6.0\arcsec$ beam, so the source is unresolved at $6\arcsec$ ($28\,\mathrm{kpc}$). }

\section{Refractive interstellar scintillation and compactness}
\label{app:riss}

We assess whether refractive interstellar scintillation (RISS) could contribute to the tentative 3-GHz variability. For the sightline to J092359+415736, NE2001 gives $\mathrm{SM}=6.5\times10^{-5}\,\mathrm{kpc\,m^{-20/3}}$ \citep{2002astro.ph..7156C}. With a representative $1$-kpc scattering screen, the prescription of \citet{1998MNRAS.294..307W} gives a transition frequency of $\nu_0\simeq6.16$ GHz. The 3-GHz observations are therefore in the strong-scattering regime. Their refractive timescale is $\simeq0.55$ d for $v_\perp=50\,\mathrm{km\,s^{-1}}$, so survey snapshots separated by years sample the process stochastically and cannot exclude a RISS contribution.

To scintillate with the point-source modulation index at 3 GHz, the emitting component must satisfy $\theta_s\lesssim15.9\,\mu$as \citep{1998MNRAS.294..307W}. If it contains the weighted mean flux density of $408\,\uJy$, the corresponding brightness temperature is
\begin{equation}
  T_{\rm b} = \frac{2\ln 2}{\pi k_{\rm B}}\,\frac{S_\nu\,\lambda^2}{\theta_s^2}
    \gtrsim 2.2\times10^{11}\,\mathrm{K}\quad(\text{observer frame}).
\end{equation}
At $z=0.323$, this gives $T_{\rm b,rest}\gtrsim2.9\times10^{11}$ K. The numerical value depends on the tentative variability and the uncertain scattering model. If only a fraction $f$ of the flux density scintillates, the lower bound scales as $f$. Nevertheless, RISS can contribute appreciably only if the radio emission includes an ultracompact nuclear component. This is incompatible with extended star formation, whose brightness temperatures are typically $\lesssim10^5$ K \citep{1992ARA&A..30..575C}.

\section{Error budget and treatment of low-significance epochs}
\label{app:errors}

\paragraph{\rev{Position offset and peak-fitting tests.}}
\label{app:offsetfit}

We fitted the radio emission both at the \textit{Gaia} DR3 position and at the local maximum within $0.5$ of the beam minor axis. For ASKAP detections above $3\sigma$, the fitted amplitudes differ by $\lesssim8\%$. At $\sim2\sigma$, the local maximum becomes sensitive to nearby noise fluctuations, and we use the flux density measured at the \textit{Gaia} position. FIRST and VLASS residual maps after subtraction of \textit{Gaia}-centred models show no recurring compact component at a common offset.

We report the raw image-plane amplitude of the beam-fixed FIRST fit without applying the survey-standard $+0.25\,\mathrm{mJy}$ CLEAN-bias correction \citep{2015ApJ...801...26H}, which was calibrated for catalogue detection and cannot be verified independently at low S/N from the delivered image alone. For FIRST, we injected sources with amplitudes of $0.25$--$1.0 \,\mathrm{mJy}$ at 500 blank positions in the delivered image. The recovered amplitudes have a mean offset below $5 \,\uJy$ and a scatter of $0.13$--$0.14 \,\mathrm{mJy}$. At the survey angular resolutions, nuclear emission cannot be separated spatially from unresolved host-galaxy emission.

\paragraph{Internal consistency of the RACS-mid measurement.}
\label{app:racsinternal}

RACS-mid gives $1276\pm257\,\uJy$ at 1.367 GHz. The RACS-mid and
RACS-high images, separated by 357 d, each recover a
$\sim1\,\mathrm{mJy\,beam^{-1}}$ peak at the \textit{Gaia} DR3
position at $\geq3.9\sigma$. Aperture integration of RACS-mid within
$r<0.75\,\theta_{\rm maj}$ gives $\simeq1.0$ mJy, close to the
fitted peak intensity.

Positional and peak-fitting tests are described in
Appendix~\ref{app:offsetfit}. Searches of a
$60\arcsec\times60\arcsec$ field after masking the target find no
companion above four times the local rms. The largest off-source
excursions are $3.2$--$3.6\sigma$. At the available angular
resolution, nuclear emission remains unresolved from host-galaxy
emission.

FIRST gives $428\pm139\,\uJy$ at 1.435 GHz, compared with
$1276\pm257\,\uJy$ in RACS-mid at 1.367 GHz. The LoTSS DR3 image
has a $6.0\arcsec$ beam, matching FIRST, and recovers
$2234\pm235\,\uJy$ at 144 MHz. The LoTSS--ASKAP spectrum predicts
$1121\,\uJy$ at 1.435 GHz, so FIRST remains a factor of $\sim2.6$
fainter at matched resolution. Searches of the LoTSS field find no
compact companion above $4\sigma$ within $60\arcsec$. The FIRST
deficit is therefore not explained by missing extended emission
between $5\arcsec$ and the ASKAP beam. The 26 yr contrast remains
the main long-baseline variability evidence.

\paragraph{Low-significance epochs.}

VLASS 1.2 and VLASS 2.2 have $S_\nu/\sigma_{\rm tot}<3$, with fitted flux densities of $357\pm139 \,\uJy$ and $257\pm129 \,\uJy$. Their corresponding $3\sigma_{\rm tot}$ noise levels are $417 \,\uJy$ and $387 \,\uJy$. Both epochs enter the three-point 3-GHz light curve as measured flux densities.

The LoTSS DR3 detection at $42\sigma$ establishes the radio counterpart, and seven GHz measurements above $3\sigma$ recover it over three decades. The FIRST contrast relative to the LoTSS--ASKAP spectrum is given in Appendix~\ref{app:racsinternal}. \rev{VAST-mid 1 is included as a measured flux density rather than an upper limit.}

\section{Spectral coverage at high redshift.}
\label{app:highz}

At $z=5$ the observed 0.144--3\,GHz band maps to rest-frame 0.86--18\,GHz, a factor of $\sim6$ higher than the rest-frame frequencies probed at $z=0.323$. The single power law measured locally therefore samples a different part of the intrinsic spectrum from that accessible to the same surveys at high redshift. In particular, synchrotron self-absorption or free--free absorption turnovers that fall below $\sim1\,\mathrm{GHz}$ in the rest frame would be shifted into the survey bands for sources at $z\gtrsim4$, potentially suppressing the observed flux density even when the intrinsic luminosity is comparable to that of J092359+415736. 
Moreover, a recent model invokes high-inclination, hyper-Eddington LRDs whose self-shielding can make their inferred Eddington ratios appear low \citep{2026arXiv260621105Z}. This provides one possible reason why their radio properties need not track those of low-redshift analogues.

\end{appendix}

\clearpage
\end{document}